# Temperature Dependence of *β*-Ga$_2$O$_3$ Heteroepitaxy on c-plane Sapphire using Low Pressure Chemical Vapor Deposition

Gavax Joshi[*], Yogesh Singh Chauhan, and Amit Verma[#]

Department of Electrical Engineering, Indian Institute of Technology Kanpur, India
Email: [*]gbjoshi@iitk.ac.in, [#]amitkver@iitk.ac.in

**Abstract.** *β*-Ga$_2$O$_3$ has drawn significant attention for power electronics and deep ultraviolet (UV) photodetectors owing to its wide bandgap of ~ 4.4 - 4.9 eV and high electric breakdown strength ~7-8 MV/cm. Growth of *β*-Ga$_2$O$_3$ epitaxial thin films with high growth rate has been recently reported using low pressure chemical vapor deposition (LPCVD) technique. In this work, we have investigated the effect of growth temperature on *β*-Ga$_2$O$_3$ films grown on c-plane sapphire substrates using LPCVD. We performed growths by varying temperatures from 800 °C to 950 °C while keeping all other growth parameters (Ar/O$_2$ gas flow rates, growth pressure, and Gallium precursor to substrate distance) constant. Optical, structural, and surface characterizations are performed to determine the bandgap, phase purity, crystal orientation, and crystalline quality of the grown thin films. Amorphous islands of Ga$_2$O$_3$ are observed at growth temperature of 800 °C while continuous and crystalline (-201) oriented *β*-Ga$_2$O$_3$ thin films are achieved for growth temperatures of 850 °C to 950 °C. Crystallinity of the films is found to improve with increase in growth temperature with a minimum rocking full width at half maximum of 1.52° in sample grown at 925 °C. For all the samples grown at and above 875 °C, transmittance measurements revealed an optical bandgap of ~4.77-4.80 eV with high growth rate of ~6 μm/hr.

**Keywords:** *β*-Ga$_2$O$_3$, low pressure chemical vapor deposition (LPCVD), wide bandgap, heteroepitaxy, thin film.

## 1. INTRODUCTION

Gallium Oxide (Ga$_2$O$_3$) crystallizes in five polymorphic forms: $\alpha$, $\beta$, $\gamma$, $\delta$, and $\epsilon$ [1]. Amo[ng these polymorphs, *β*-Ga$_2$O$_3$ is the most thermodynamically stable polymorph at room temperature [1]. *β*-Ga$_2$O$_3$ has a monoclinic crystal structure with a wide energy band gap of ~ 4.4 - 4.9 eV and high electrical breakdown ($E_{br}$) strength of ~ 7-8 MV/cm [2-4]. This theoretical estimation of $E_{br}$ is ~3 times larger than other wide bandgap semiconductors such as GaN and SiC [5]. High $E_{br}$ results in higher Baliga's and Johnson's figure of merit for *β*-Ga$_2$O$_3$ indicating its potential for highly efficient power devices with reduced conduction losses [6]. In the last few years, *β*-Ga$_2$O$_3$ Schottky barrier diodes with fast switching and high breakdown voltage [6-7], and *β*-Ga$_2$O$_3$ field-effect transistors (FETs) with lateral/vertical device structure [8-10], have been demonstrated with high power figure of merit and large breakdown voltage. Due to its wide bandgap, *β*-Ga$_2$O$_3$ has found applications as high performance solar-blind ultra-violet (UV)-photodetectors as well [11,12]. Properties of *β*-Ga$_2$O$_3$ are also suitable to realize gas sensors for harsh environment applications [13,14]. Because of its high laser damage threshold and high optical power tolerance, *β*-Ga$_2$O$_3$ is also a promising candidate for realizing dielectric laser accelerators and low loss plasmonics [15]. Besides its basic material advantages for various applications, β-Ga$_2$O$_3$ also has an economical advantage over other wide-bandgap semiconductors because of the availability of high quality, large area single crystalline β-Ga$_2$O$_3$ substrates produced using low cost melt growth methods [16].

In order to realize various applications of β-Ga$_2$O$_3$, high quality epitaxial thin films of β-Ga$_2$O$_3$ with controlled doping are needed. Power electronics applications specially need thick vertical epilayers with low doping [17]. Epitaxial growth of β-Ga$_2$O$_3$ has been investigated using molecular beam epitaxy (MBE) [18-19], metal organic chemical vapor deposition (MOCVD) [20,21], mist-CVD [22], halide vapor phase epitaxy [23], pulse laser deposition [24], atomic layer deposition [25], and low pressure chemical vapor deposition (LPCVD) [26-29] among other techniques. LPCVD is an epitaxial growth method which has demonstrated fast β-Ga$_2$O$_3$ growth rates of up to 30 µm/hr, under specific conditions, with controlled doping [30]. LPCVD is therefore quite promising for realization of high-power semiconductor devices with high breakdown voltage, as these devices need thick epilayers with low-doping concentrations.

The growth temperature plays a crucial role in deciding film quality during epitaxial growth as it provides surface migration energy to adatoms to reach the active growth sites [31]. Its role is even more important in β-Ga$_2$O$_3$ LPCVD as the growth temperature also decides the Ga flux. Thus, a detailed investigation of the material quality of β-Ga$_2$O$_3$ thin films as a function of growth temperature is required using LPCVD. In this work, on a customised in-house build LPCVD system, we have investigated the effect of varying the Ga source and substrate temperature on grown β-Ga$_2$O$_3$ thin films over c-plane sapphire (c-Al$_2$O$_3$) substrates while keeping the other growth parameters constant. The grown films have been characterized using optical, structural and surface characterization techniques to investigate the effect of growth temperature on material quality. The details of the experiment and characterizations are presented in the following sections.

## 2. EXPERIMENTAL DETAILS

For performing growths of β-Ga$_2$O$_3$ thin films, we used a customized single zone horizontal LPCVD reactor whose schematic is shown in Fig. 1(a). High purity Gallium metal (UMC, 99.99999% purity) and O$_2$ (99.9% purity) gas were used as source precursors and Argon (99.999% purity) was used as a carrier gas. The films were grown by keeping both the Ga source material and the c-plane Sapphire substrate in the center of a quartz tube (100 cm length and 2.7 cm inner diameter) with ~5 cm separation. The typical furnace temperature, Ar and O$_2$ flow rate profiles used for growth are shown in Fig. 1(b). The growth is started by pumping down the entire system to base pressure. After 5 minutes of pumping, Ar flow of 100 sccm is started which remains on during the rest of the growth and subsequent cool down. Furnace temperature is then ramped to reach the desired growth temperature with a ramp rate of ~4 °C /min. Once desired temperature is reached and stabilized, growth is started by flowing O$_2$ at a rate of 5 sccm. The tube pressure during the growth is maintained at ~ 100 Pa. After a growth period of 30 minutes, the oxygen flow is stopped, and the furnace is switched off and left to cool naturally under Ar flow. The deposited thin film was taken out at room temperature for further characterization. Before every growth, the c-plane Sapphire substrates were cleaned using acetone, isopropanol and de-ionized water for 10 minutes in ultrasonicator bath and dried with nitrogen flow. To study the effect of growth temperature on the Ga$_2$O$_3$ thin film quality, six samples were grown with varying growth temperatures, T$_g$ = 800 °C, 850 °C, 875 °C, 900 °C, 925 °C, and 950 °C.

The deposited *β*-Ga$_2$O$_3$ thin films were characterized for optical properties, thickness, surface morphology, phase purity, and crystal orientation. Cary 7000 Universal measurement spectrophotometer (Agilent technologies) was used to take transmission spectrum in UV and Visible range (200 nm – 800 nm). To extract the thickness of *β*-Ga$_2$O$_3$ thin films on sapphire, reflectance measurement was performed on Filmetrics, F20-EXR thin film analyzer. To obtain surface roughness of the samples, atomic force microscopy (AFM) was performed using Asylum research MFP-3D infinity system. To get surface topographical information, high resolution field emission scanning electron microscope (FESEM) imaging was done on Carl Zeiss Nova nanoSEM-450 and JEOL JSM-7100F, microscope. Raman spectroscopy was carried out on Princeton Instruments Acton Spectra Pro 2500i system with 532 nm solid state laser diode excitation. In-plane 2θ and rocking scans were performed on a PANanlytical empyrean X-ray diffractometer with copper K$_α$ radiation to determine film crystal orientation and quality. All the characterization measurements were done at room temperature.

## 3. RESULTS AND DISCUSSION

### 3.1 Optical Characterization

External optical transmission spectra as a function of wavelength (200 nm to 800 nm) of all the grown thin films are shown in Fig. 2(a). For thin films grown at T$_g$ = 850 °C and above, absorption edge is observed in the UVC region around wavelength of 260 nm, suggesting growth of *β*- phase of Ga$_2$O$_3$. All the films show high optical transmission of ~60-80% in near UV and visible wavelength region. The transmittance increases with deposition temperature from ~65% at 850 °C to ~80% at 900 °C and then decreases to ~60% at 925 °C and 950 °C. The sharpness of the absorption edges also indicates crystalline nature of the *β*-Ga$_2$O$_3$ suggesting high quality thin films. The increase in transmittance is due to the improvement in crystallinity quality of β-Ga$_2$O$_3$ thin films with growth temperature and the decrease in the transmittance to ~60% at 925 °C and 950 °C, is possibly because of increase in O-vacancies in the films due to Ga-rich (O-poor) conditions at higher temperature which results in increase in absorption [32]. As *β*-Ga$_2$O$_3$ is close to a direct bandgap semiconductor [33], the power law dependence of absorption coefficient *α* is given by, $(\alpha h\nu)^2 \propto (h\nu - E_g)^{0.5}$, where $h\nu$ is the energy of the incident photon, and $E_g$ is the band gap. Tauc plots to extract the optical bandgap of the grown thin films are shown in Fig. 2(b). We find extracted optical bandgap of ~4.77-4.88 eV which agrees well with the bandgap of *β*-Ga$_2$O$_3$ [2,3,33]. Optical bandgap dependence on T$_g$ is shown in Fig. 2(c).

To measure the thickness of the films, we performed reflectance measurements. The thickness was extracted by modeling the reflectance data assuming reported values of *β*-Ga$_2$O$_3$ refractive index [34]. The extracted thickness as a function of T$_g$ is shown in Fig. 2 (c). T$_g$ = 850 °C sample was ~ 0.41 µm thick while all samples with T$_g$ ≥ 875 °C had a thickness close to ~3 µm. The smaller thickness of T$_g$ = 850 °C sample can be due to low Ga flux at this temperature. For all samples with T$_g$ ≥ 875 °C, we found that the starting ~1 g Ga precursor was completely consumed by the end of the growth. This might be the reason for same thickness in all of these samples resulting from complete consumption of the Ga precursor. The observed growth rate of ~ 6 µm/hr for these samples is actually a lower bound on the real growth rate as the specific growth time when the Ga precursor is exhausted is not known. The actual growth rate is expected to be higher than ~ 6 µm/hr for all samples with $T_g \geq 875$ °C. The observed growth rate is comparable to reported $\beta$-$Ga_2O_3$ growth rates using LPCVD [30].

The film grown at $T_g = 800$ °C shows weak absorption for wavelengths less than ~300 nm with more than ~60% transparency at 200 nm. This observation suggests growth of defective $Ga_2O_3$ which has either incomplete coverage on c-plane (0001) sapphire substrate and/or its thickness is extremely thin allowing only partial absorption. To investigate on the substrate coverage of the film, we performed AFM of the sample. The AFM (Fig. 2(d)) shows small islands (height ~ 0-100 nm) which do not coalesce to form a complete film. Similar observation can be made from Fig. 2(e) and Fig 2(f), showing surface FESEM image and EDS spectra of the sample grown at $T_g = 800$ °C. Small islands of width ~50 – 70 nm are clearly visible in FESEM. EDS composition analysis of surface shows O, Al and Ga atoms with atomic weight 53.39%, 44.02% and 2.59%, respectively. The presence of high concentration of Al confirms incomplete coverage of thin film over the entire substrate. The scattered and isolated islands of $Ga_2O_3$ are the reason for finite transmission even in deep UV for sample grown at 800 °C. This morphology is possibly because of low Ga flux and low adatom energies at temperature of 800 ℃ which leads to island growth.

### 3.2 Raman Characterization

Fig. 3 shows measured Raman spectra of grown thin films excited using a 532 nm laser. Raman peak positions of the phonon modes, as observed in samples grown at different temperatures, are listed in table 1. Raman active modes are divided into three groups: high frequency (~770-500 $cm^{-1}$) due to stretching and bending of $GaO_4$ tetrahedra, mid frequency (~480 – 310 $cm^{-1}$) assigned to deformation of $Ga_2O_6$ and low frequency modes (below 200 $cm^{-1}$) are attributed to tetrahedra-octahedra chains [35]. As clear from Table 1, the obtained Raman modes are in good agreement with experimental and theoretical values reported in the literature [36]. No Raman peaks are obtained for $Ga_2O_3$ islands obtained at $T_g = 800$ °C suggesting that the islands are amorphous. For sample grown at 850 °C, only few Raman modes are activated suggesting that the thin film is of lower crystallinity as compared to other samples. For higher $T_g$ ($\geq 875$ °C), all $\beta$-$Ga_2O_3$ phonon modes are observed in the measured Raman spectra which clearly indicates growth of phase pure $\beta$-$Ga_2O_3$ thin films.

### 3.3 XRD Characterization

XRD patterns of as-grown $Ga_2O_3$ thin films at different substrate temperatures are shown in Fig. 4. For $T_g = 800$ °C, no film diffraction peak is observed suggesting growth of amorphous $Ga_2O_3$. This observation is consistent with the Raman measurements discussed in previous subsection. For thin films grown at $T_g \geq 850$ °C, three XRD diffraction peaks are observed at ~18.9°, ~38.3° and ~58.9° respectively for all the samples. This indicates that the deposited $Ga_2O_3$ thin films are of (-201) orientation associated with the $\beta$-phase of $Ga_2O_3$. The (-201) oriented family of planes is energetically more favorable over c-plane sapphire, because atomic configuration of the (-201) plane matches with that of c-plane sapphire [37]. For $T_g = 875$ °C, a small diffraction peak is also observed at 48.7°, which corresponds to (510) peak of $\beta$-Ga$_2$O$_3$. This peak vanishes in samples grown at higher growth temperatures.

From the above discussion, we can conclude that all the samples with T$_g$ ≥ 850 °C show single phase (-201) oriented $\beta$-Ga$_2$O$_3$ thin films. However, the ratio of the intensity of (-201) and (-402) diffraction peaks (I$_{(-201)}$/I$_{(-402)}$) and the ratio of the intensity of (-603) and (-402) diffraction peaks (I$_{(-603)}$/I$_{(-402)}$) shows variation as T$_g$ increases from 850 °C to 950 °C as shown in Fig. 5(a). Since there are no film diffraction peaks observed for sample grown at 800 °C, its intensity ratios are not shown in Fig. 5(a). The ideal values for these intensity ratios for $\beta$-Ga$_2$O$_3$ are also marked in the figure [38]. With increasing growth temperature, we observe that the measured XRD intensity ratios approach ideal values suggesting an improvement in the crystalline quality of the thin films. Similar conclusions can be drawn from Fig. 5(b), which shows decrease in full width half maximum (FWHM) of (-201), (-402) and (-603) diffraction peaks with increasing T$_g$. FWHM is calculated by fitting a gaussian function to each XRD peak. FWHM values decrease with increase in growth temperature, suggesting improvement in crystallinity of $\beta$-Ga$_2$O$_3$ thin films. Crystallite size is calculated using Scherrer (Eq. 1) [39] and Williamson-Hall (W-H) (Eq. 2) [39] formula where $k$ is Scherrer's constant = 0.94, $\lambda$ is wavelength of the X-ray radiation = 0.154060 nm, θ is the Bragg's angle, η is strain in the material, $B_r$ is the FWHM of the diffraction peak, and $D_S$ and $D_{WH}$ are calculated crystallite size using Scherrer's and W-H method, respectively. Lower angle diffraction peak, (-201) reflection is considered for $D_S$ calculation [39]. $D_{WH}$ and strain is calculated from the intercept and the slope, respectively, by plotting $B_r cos\theta$ vs $sin\theta$ from Eq. 2 and fitting the data linearly.

$$D_S = \frac{k\lambda}{B_r \cos\theta} \qquad Eq.\ 1$$

$$B_r \cos\theta = \frac{k\lambda}{D_{WH}} + \eta \sin\theta \qquad Eq.\ 2$$

A plot comparing $D_S$ and $D_{WH}$ as a function of growth temperature is shown in Fig. 5(c). $D_S$ increases from ~25 nm at T$_g$ = 850 °C to ~27 nm for T$_g$ ≥ 875 °C. In comparison, $D_{WH}$ is constant at ~28 nm and does not change with T$_g$; the films are however found to be slightly strained with the strain decreasing with increasing growth temperature. This suggests that the $\beta$-Ga$_2$O$_3$ thin films deposited at higher temperature are much more relaxed. These results are consistent with previous $\beta$-Ga$_2$O$_3$ growth reports using different epitaxial methods [40-42].

Fig. 5(d) shows the XRD rocking scan of (-402) family of planes for $\beta$-Ga$_2$O$_3$ films grown at 850 °C and above. The rocking FWHM values of the films grown at 850 °C, 875 °C, 900 °C, 925 °C and 950 °C are 1.61°, 2.12°, 1.91°, 1.5°, and 1.64°, respectively, as shown in Fig. 5(e). Among the samples of similar thickness (T$_g$ = 875 °C - 950 °C), the rocking FWHM value is highest for $\beta$-Ga$_2$O$_3$ film grown at 875 °C and lowest for film grown at 925 °C. This decrease in FWHM value of the diffraction peak suggests improvement in the crystal quality with growth temperature till T$_g$ = 925 °C. To explore the nature of epitaxial relationship between $\beta$-Ga$_2$O$_3$ thin film and c-plane sapphire substrate, we performed off-axis ϕ scan for diffraction plane (111) for T$_g$ = 925 °C sample (Fig. 5(f)). In-plane 6-fold rotational symmetry is visible from presence of six peaks separated by 60°. Since, $\beta$-Ga$_2$O$_3$ has a 2-fold in-plane symmetry which is grown on a c-plane sapphire substrate with 3-fold rotational symmetry, the presence of 6 peaks in the ϕ scan suggests presence of rotational domains in the thin film [43].

### 3.4 Surface Characterization

Fig. 6 shows top surface AFM scan for the scan area of 5 μm x 5 μm for samples deposited at different temperatures. The dependence of surface morphology, and crystallinity on the growth temperature is clearly visible in images. β-$Ga_2O_3$ islands formed at 800 °C (Fig. 2(d)) gets coalesced into a continuous thin film at higher growth temperature. As deposition temperature is increased (≥ 850 °C) the crystallinity is improved and change in surface morphology is also observed. As shown in Fig. 6 graph, variation in mean RMS roughness is seen as growth temperature is increased from 850 °C to 900 °C and stays almost constant for higher growth temperatures. The smoothest thin films are obtained at $T_g$ = 850 °C, which is also confirmed by interference fringes in transmittance spectra as shown in Fig. 2(a).

To analyse the top surface morphology and elemental composition, FESEM and EDS was performed. Top view FESEM image of $T_g$ = 950 °C sample is shown in Fig. 7(a). Hexagonal crystal edges at the surface are clearly visible which matches with the 6-fold symmetry of the substrate and the (-201) *β*-$Ga_2O_3$. EDS spectra of the sample was done to determine the chemical composition. EDS spectra result is shown in Fig. 7(b). The atomic weight percentage of Ga and O was found to be ~77.9% and ~22.1%, respectively, suggesting that the grown films have the correct stoichiometric composition.

## 4. CONCLUSIONS

*β*-$Ga_2O_3$ thin films were synthesized on c-plane sapphire substrate using LPCVD. The effect of varying the growth temperature (from 800 °C to 950 °C) on the optical and the structural properties of the *β*-$Ga_2O_3$ films was studied. Amorphous $Ga_2O_3$ islands were obtained for the growth temperature of 800 °C, while continuous and crystalline (-201) *β*-$Ga_2O_3$ films were obtained for the growth temperature of 850 °C - 950 °C. Films grow with a high growth rate of ~6 μm/hr for $T_g$ of 875 °C - 950 °C. Crystallinity of the films was found to significantly improve with the increase in growth temperature with the best quality films obtained for $T_g$ = 900 °C - 950 °C. LPCVD growth of *β*-$Ga_2O_3$ thin films with good crystalline quality at high growth rates will help enable the applications of this material in high power transistors, deep UV photodetectors, and harsh environment sensors.

## ACKNOWLEDGEMENT

This work was supported by SERB Early Career Research Award (Grant No. ECR/2018/001076) and used XRD (supported by DST-FIST Grant No. SR/FST/ETII-053/2012) and Raman characterization facilities of Dept. of Materials Science and Engineering, IIT Kanpur. Authors also thank A. Kalra and Dr. D. Nath from Indian Institute of Science, Bangalore for help with few XRD measurements.

Table 1: Measured Raman peak spectral positions of the phonon modes in LPCVD grown $\beta$-Ga$_2$O$_3$ thin films

(n.o = not observed).

| Raman phonon modes | Growth Temperature | | | | | Experimental Reference [36] | Theoretical Reference [36] |
|---|---|---|---|---|---|---|---|
| | 850 °C | 875 °C | 900 °C | 925 °C | 950 °C | | |
| $B_g^2$ | 143.3 | 146.6 | 146.6 | 146.6 | 145.2 | 144.8 | 145.6 |
| $A_g^2$ | 168.9 | 169.5 | 169.4 | 170.8 | 170.8 | 169.9 | 176.4 |
| $A_g^3$ | 199.7 | 200.3 | 200.3 | 200.3 | 200.3 | 200.2 | 199.1 |
| $A_g^4$ | n.o. | 319.8 | 319.8 | 319.8 | 321.1 | 320.0 | 318.5 |
| $A_g^5$ | 348.2 | 347.4 | 347.4 | 348.7 | 347.4 | 346.6 | 342.5 |
| $A_g^6$ | 416.2 | 416.8 | 416.8 | 416.8 | 416.8 | 416.2 | 432 |
| $A_g^7$ | n.o. | 475.2 | 476.4 | 475.2 | 479 | 474.9 | 472.8 |
| $A_g^8$ | n.o. | 631.4 | 632.6 | 633.9 | 631.4 | 630.0 | 624.4 |
| $B_g^5$ | 657.4 | 652.9 | 652.9 | 655.4 | 652.9 | 652.3 | 653.9 |
| $A_g^{10}$ | 770.3 | 767.1 | 768.4 | 769.6 | 767.1 | 766.7 | 767.0 |

# List of Figures

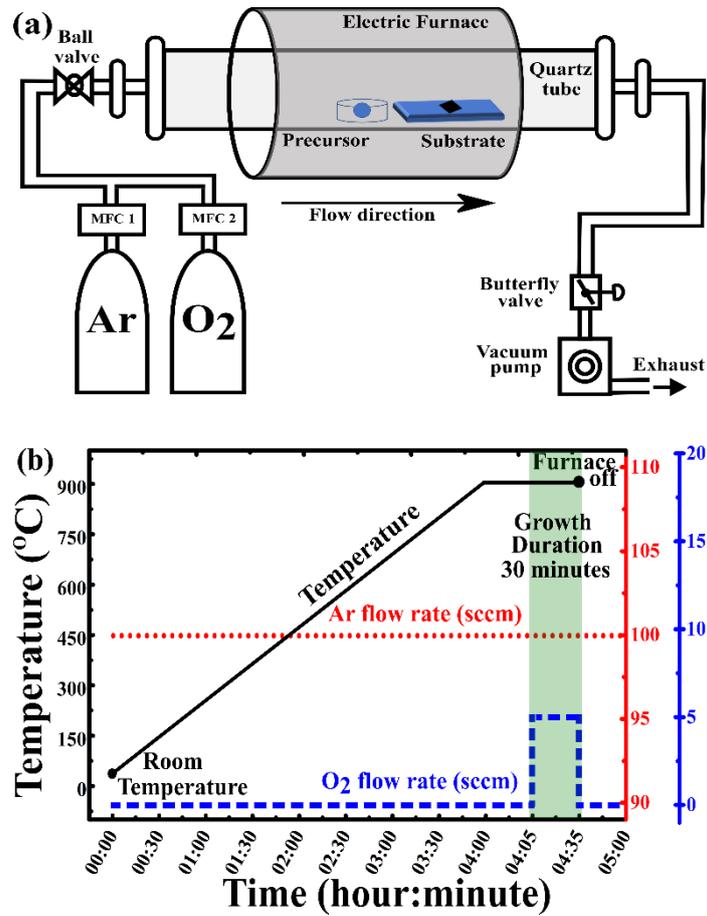

**Figure 1:** (a) Schematic of the LPCVD reactor setup used for the synthesis of *β*-Ga$_2$O$_3$ thin films, and (b) Temperature (for T$_g$ = 900 °C), Argon gas flow rate, and O$_2$ gas flow rate profiles for LPCVD growth of *β*-Ga$_2$O$_3$ thin film.

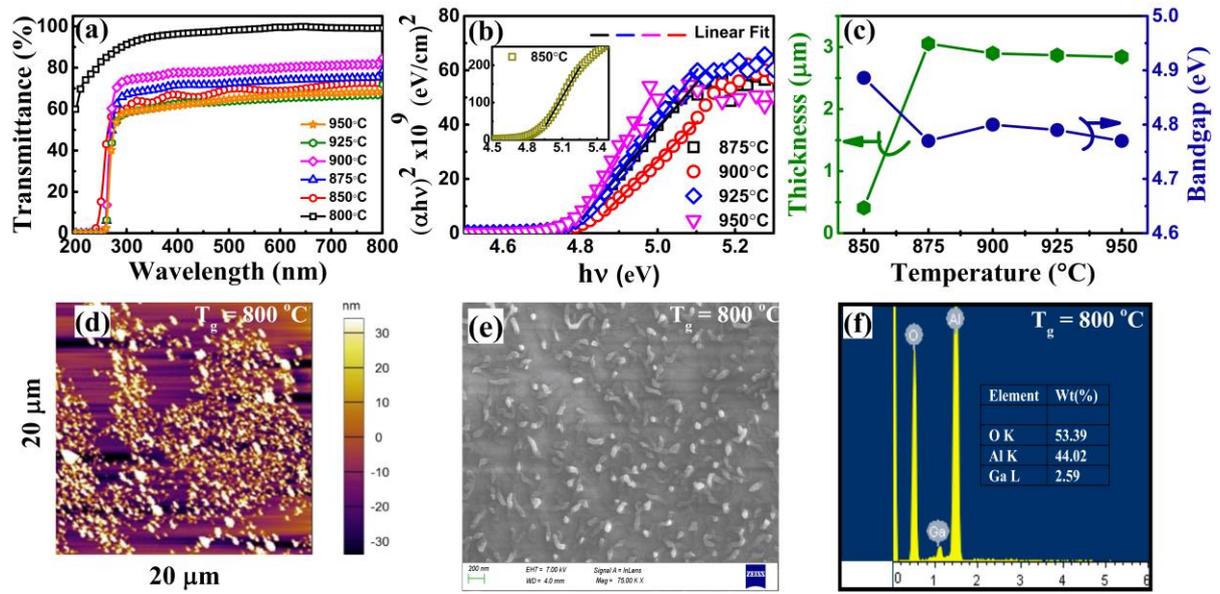

**Figure 2:** (a) UV Visible transmission spectroscopy, (b) Tauc plot of *β*-Ga$_2$O$_3$ thin films grown at different temperatures on c-plane sapphire substrate (the x- and y- axis label of inset is same as the main figure), (c) thickness (left Y axis) extracted by modelling reflectance vs wavelength data and bandgap (right Y axis) obtained from Tauc plot of *β*-Ga$_2$O$_3$ thin films with respect to the growth temperature (solid line is only given as a guide to eye), and top surface (d) AFM image (e) FESEM image (scale bar 200 nm) and, (f) EDS spectra for sample grown at deposition temperature of 800 °C.

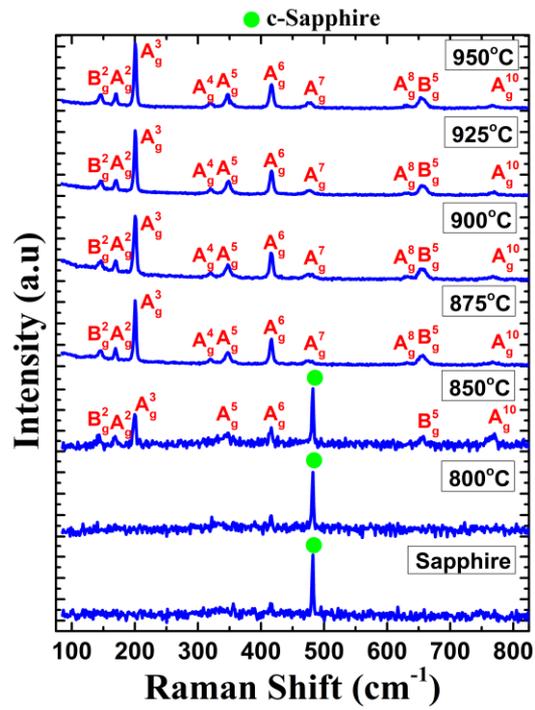

**Figure 3:** Experimental Raman Spectra of *β*-Ga$_2$O$_3$ thin films on c-plane sapphire substrates, excited by laser wavelength of λ = 532 nm, grown at different deposition temperatures.

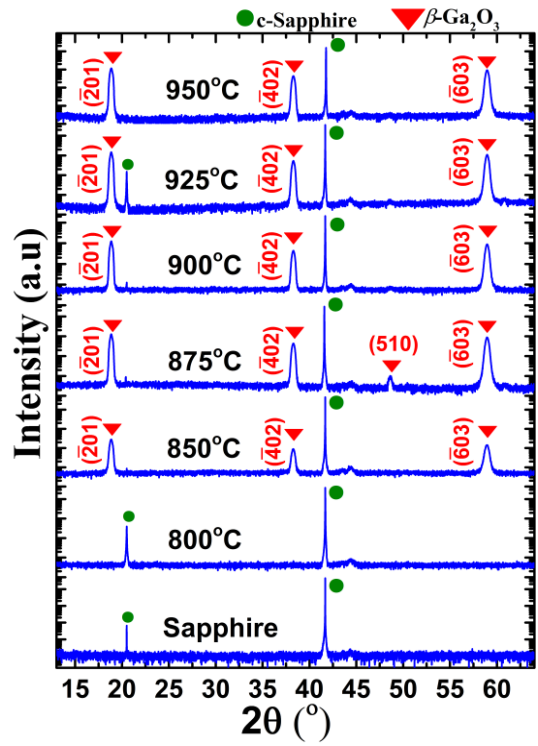

**Figure 4:** XRD spectra (ω-2θ scan) of *β*-Ga$_2$O$_3$ thin films on c-plane sapphire substrates grown at different growth temperatures.

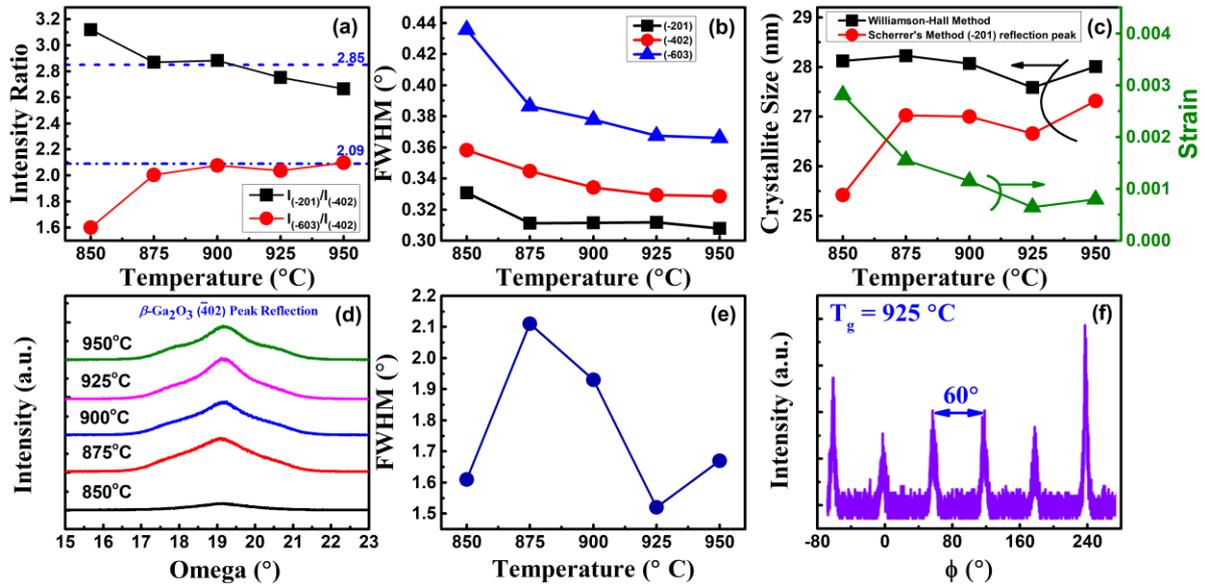

Figure 5: (a) Ratio of the intensity of (-201) and (-603) diffraction peaks to the intensity of (-402) diffraction peak (ideal intensity ratios are also marked in the figure), (b) FWHM of major peaks (-201), (-402), and (-603) from ω-2θ scan obtained by Gaussian fit, (c) On left axis: crystallite size calculated using Scherrer's and Williamson-Hall methods, On right axis: strain calculated from W-H method at different growth temperatures, (d) XRD rocking scan of (-402) diffraction peak, (e) FWHM of (-402) diffraction peak for samples grown at different temperatures, (f) In-plane XRD ϕ scan for $β$-$Ga_2O_3$ thin film grown at substrate temperature of 925 ℃ (Solid lines in Fig. 5 (a), (b), (c), and (e) are for guiding purpose only).

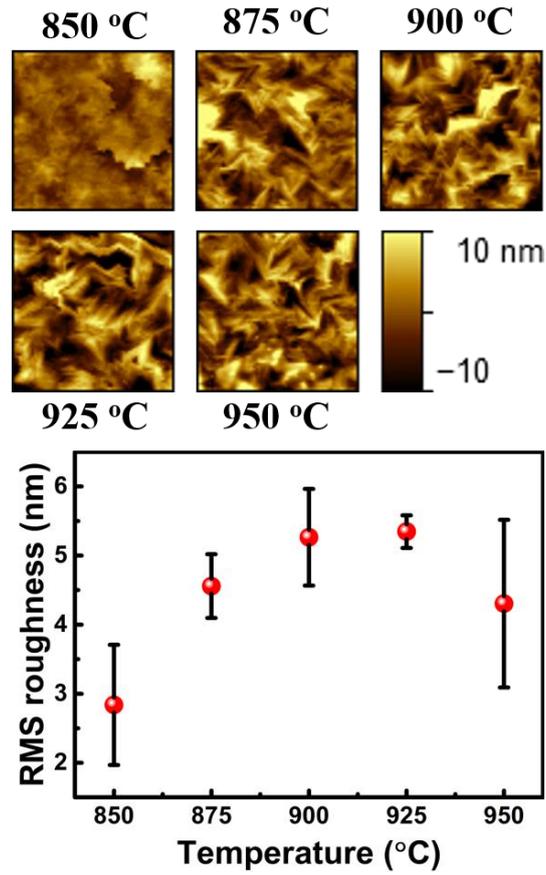

**Figure 6: AFM images for the scan area of 5 *µ*m × 5 *µ*m of *β*-Ga$_2$O$_3$ thin films deposited at different temperatures. Graph shows mean surface RMS roughness extracted from scan images as a function of deposition temperature.**

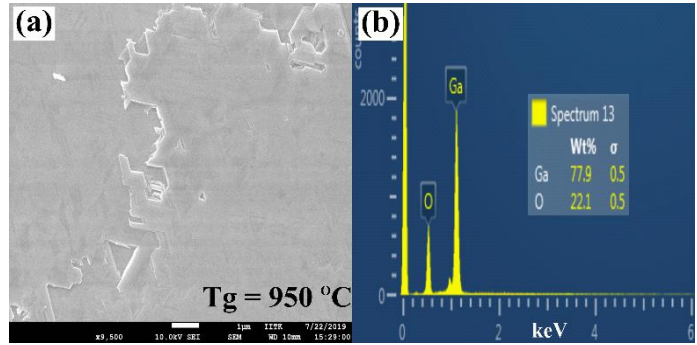

**Figure 7:** (a) Top view FESEM image of (-201) oriented *β*-$Ga_2O_3$ thin film (scale bar = 1 µm), and (b) EDS spectra of *β*-$Ga_2O_3$ thin film grown at 950 °C on c-plane (0001) sapphire substrate.